\documentclass{ab}

\usepackage{graphicx}
\usepackage{natbib}

\begin{document}

\title{New Extended Radio Sources From the NVSS}

\author{V.R.~Amirkhanyan$^1$, V.L.~Afanasiev$^2$ and A.~V.~Moiseev$^2$}

\institute{$^1$Sternberg Astronomical Institute, Moscow M.V. Lomonosov State University, Moscow, 119992 Russia\\
$^2$Special Astrophysical Observatory, Russian Academy of Sciences, Nizhnij Arkhyz, 369167, Russia}

\titlerunning{New extended radio sources from the NVSS}
\authorrunning{Amirkhanyan et al.}%

\date{October 27, 2014/Revised: November 24, 2014}
\offprints{V. Amirkhanyan  \email{amir@sao.ru}}

\abstract{
We report the results of the spectroscopic observations carried out at the SAO RAS 6-m telescope for the optical components of nine new extended radio sources found in the NVSS catalog. The measured redshifts of the host  galaxies are in the range of   $z=0.1$--$0.4$.  The physical sizes of radio sources were calculated within the standard cosmological model. The two  most extended objects, 0003+1512 and 0422+0351 reach the sizes of 2.1~Mpc and 4.0~Mpc, respectively. This is close to the maximum size of known radio sources.
}

\maketitle

\section{Introduction}

The statistics of extended radio sources required for the
cosmological tests is currently insufficient. Its expansion is only
possible  in the region of faint fluxes. The observational  foundation of this process are the  NVSS \citep{Condon} and  FIRST~\citep{White} surveys.
We investigated the NVSS catalog using the modified programs from \citet{Amirkhanian}. The result is a list of more than two thousand extended
radio source candidates. The redshift of most of them is unknown. In this work we have obtained the optical spectra and determined the redshifts of
nine radio sources from this list. Before starting the observations, we have to find the object whose spectrum we expect to study. The identification of extended
radio sources is a difficult task, since  the probability of a false identification increases together with radio source angular size. The search for the candidates in the radio sources host galaxies was carried out in several stages.
Originally, the  NED and SDSS databases were searched for the objects
(galaxies or quasars), located from the coordinates of the studied
radio sources by no more than   $30\arcsec$.  Next, we sought for the component of the radio source, closest to the optical object. And if
the distance between them was not in excess of  $10\arcsec$,  or the optical object was not further than $10\arcsec$ from the major axis
of the radio source, we  considered it the most likely optical component. This algorithm is easily formalized and gives a confident
result if the morphology of the radio source is close to classical. In more complex cases, we used the  ALADIN   software \citep{Bonnarel} to visually verify the identification.

\begin{table}[b]
\caption{Log of observations}
\label{tab1}
\medskip
\begin{tabular}{c|c|c}
\hline
 Object         &   Date of         &  Exposure,  \\[-5pt]
 name         &    observations &     s       \\
\hline
0422+1510  &   20/21.02.2012     &  $600\times2$ \\
0422+1407  &   21/22.02.2012     &  $900\times2$ \\
0638+3334  &   21/22.02.2012     &  $900\times2$ \\
1427+2657  &   12/13.05.2013     &  $600\times1$ \\
1240+5330  &   12/13.05.2013     &  $600\times1$ \\
1421+1016  &   14/15.06.2013     &  $600\times1$ \\
1301+5408  &   14/15.06.2013     &  $600\times2$ \\
1600+6137  &   14/15.06.2013     &  $300\times2$ \\
0003+0351  &   06/07.11.2013     &  $300\times3$ \\
\hline
\end{tabular}
\end{table}

\begin{figure*}
\centerline{
\includegraphics[width=7cm]{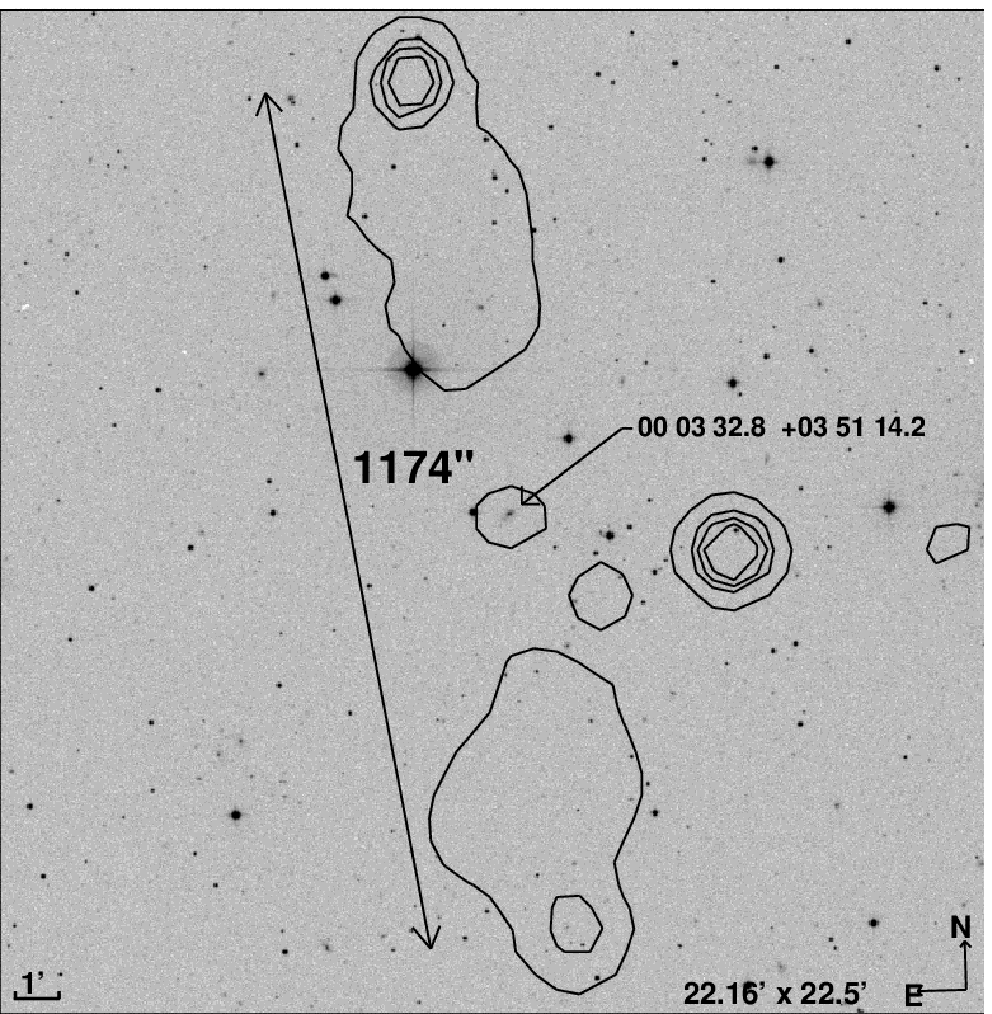}
 \hspace{1mm}
\includegraphics[width=9.4cm,bb=0 15 285 215,clip]{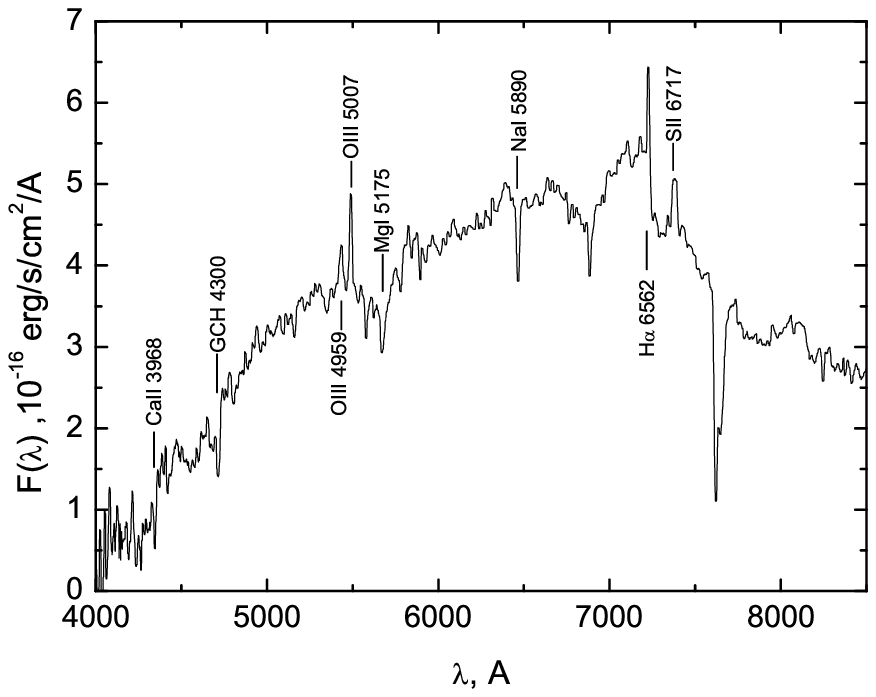}
 }
\caption{The radio source 0003+0351. Left plot: the contours of the NVSS radio map,
superimposed onto the image from the DSS2-red. The arrow marks
the position of the host galaxy and its optical coordinates,  the accepted size of the   radio structure is also denoted. Right plot: the optical spectrum of the
host galaxy:  the positions of the main emission and
absorption lines are marked.} \label{fig1}
\end{figure*}

\begin{figure*}
\centerline{
\hspace{2mm}\raisebox{9mm}{\includegraphics[width=7cm]{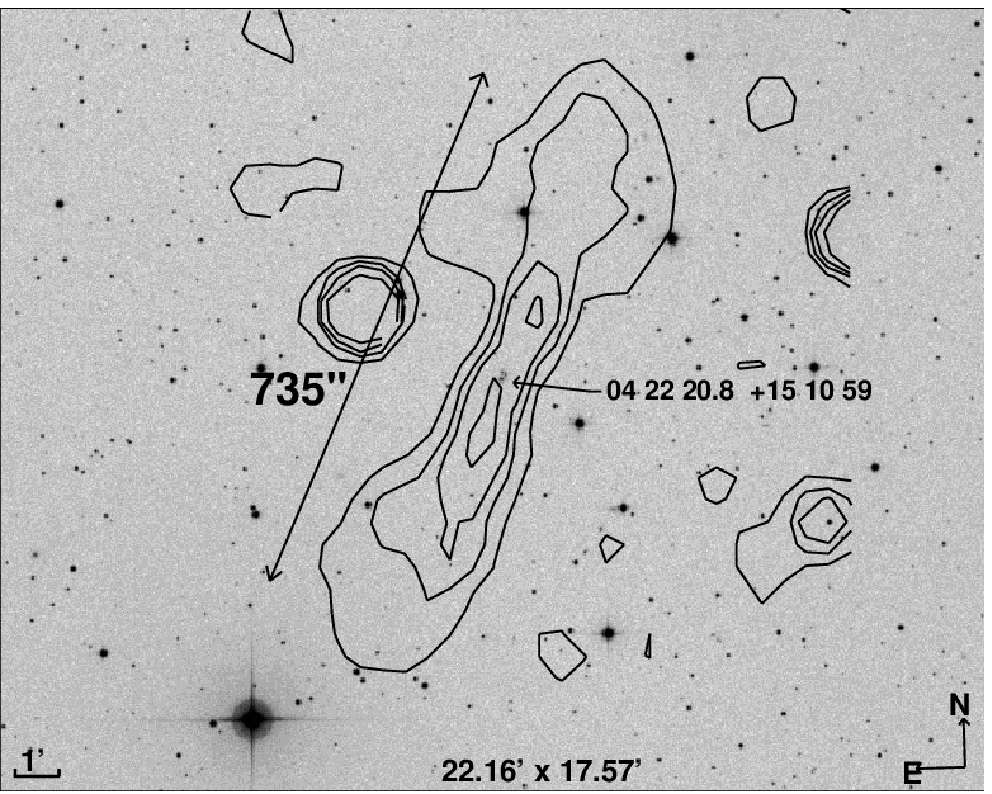}}
\includegraphics[width=9.4cm,bb=0 15 285 215,clip]{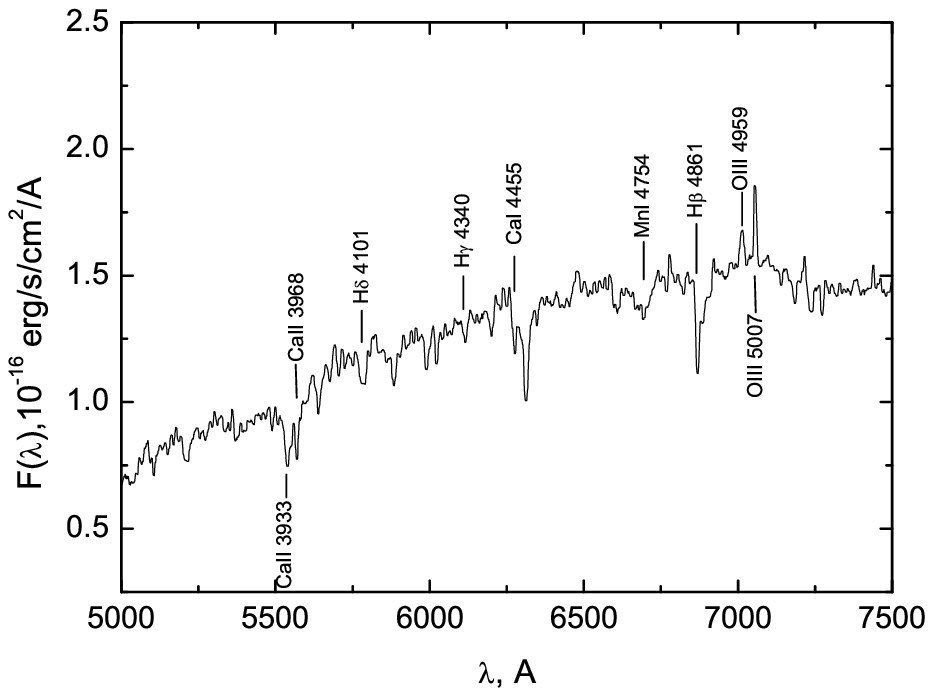}}
 \caption{Similar to
Fig.~\ref{fig1}  for the source 0422+1512. } \label{fig2}
\end{figure*}

\begin{figure*}
\centerline{
\hspace{2mm}\raisebox{9mm}{\includegraphics[width=7cm]{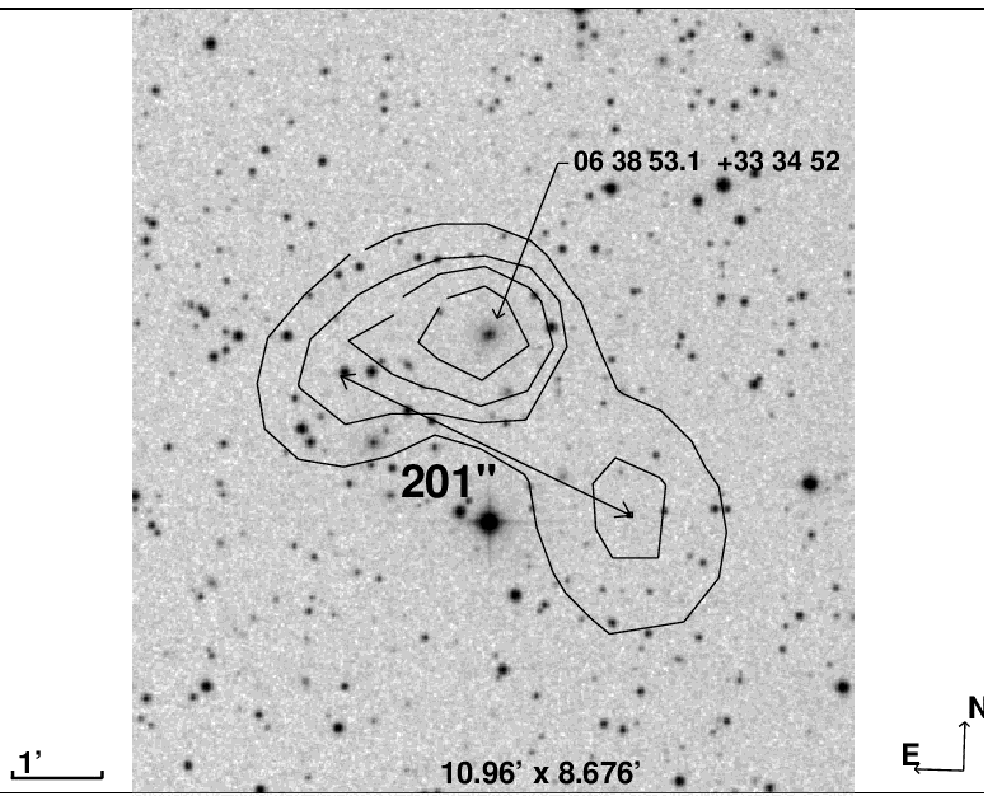}}
\includegraphics[width=9.4cm,bb=0 15 285 215,clip]{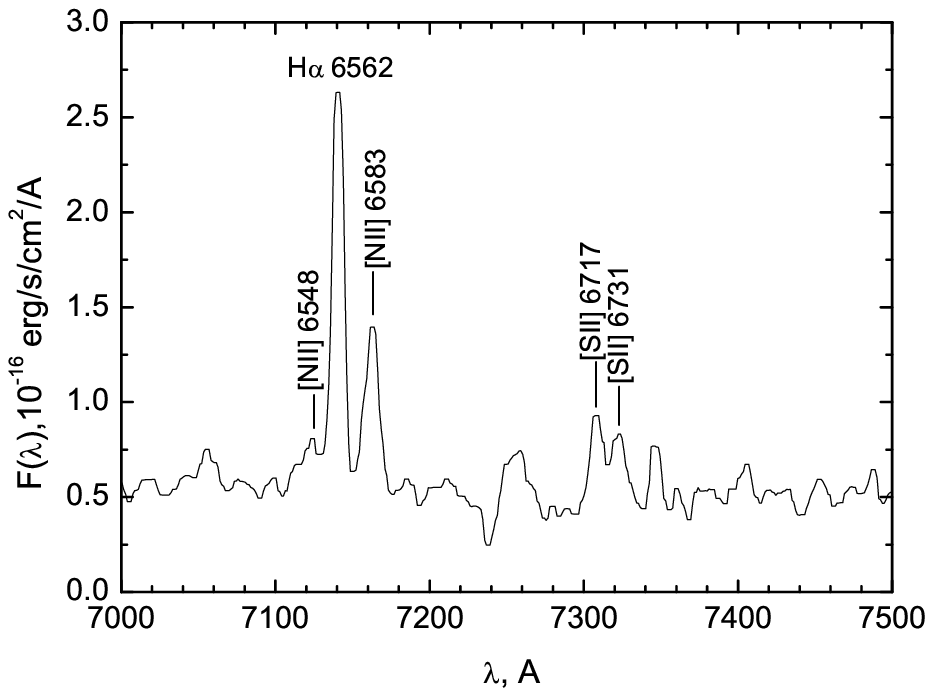}}
\caption{Similar to Fig.~\ref{fig1}  for the source
0638+3334. } \label{fig3}
\end{figure*}

\begin{figure*}
\centerline{
\hspace{5mm}\raisebox{9mm}{\includegraphics[width=7cm]{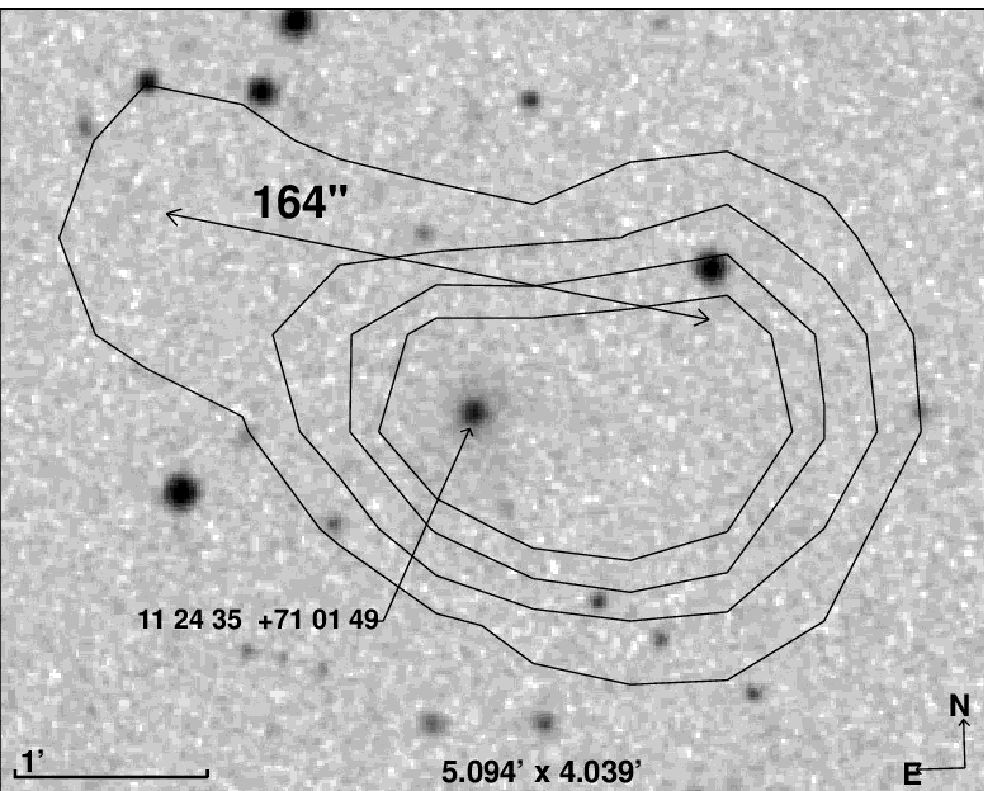}}
 \hspace{2mm}
\includegraphics[width=9.4cm,bb=0 15 285 215,clip]{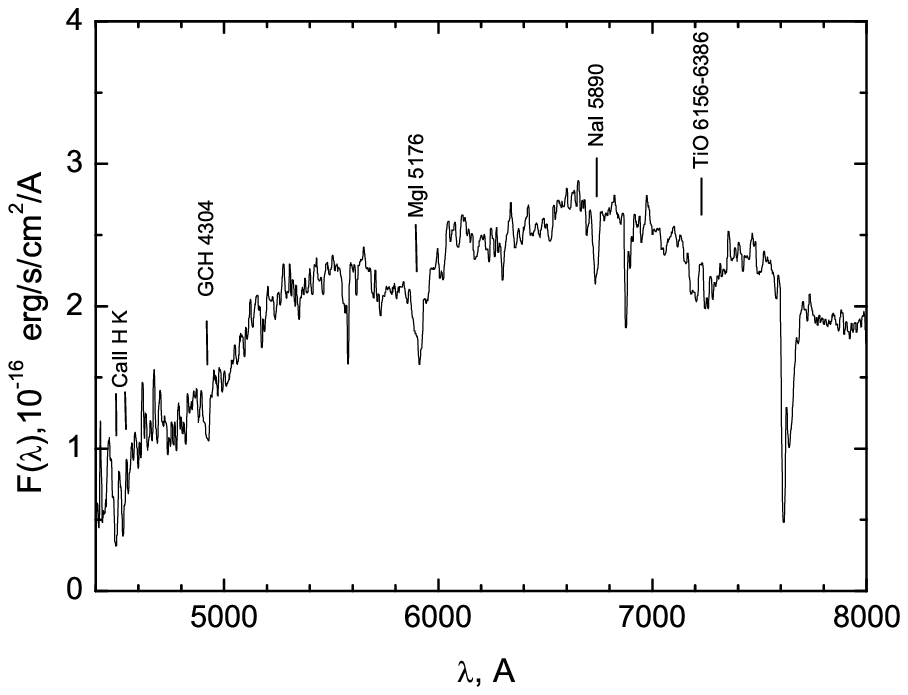}
}
\caption{Similar to Fig.~\ref{fig1}  for the source
1124+7102.} \label{fig4}
\end{figure*}

\begin{figure*}
\centerline{
\hspace{2mm}\raisebox{9mm}{\includegraphics[width=7cm]{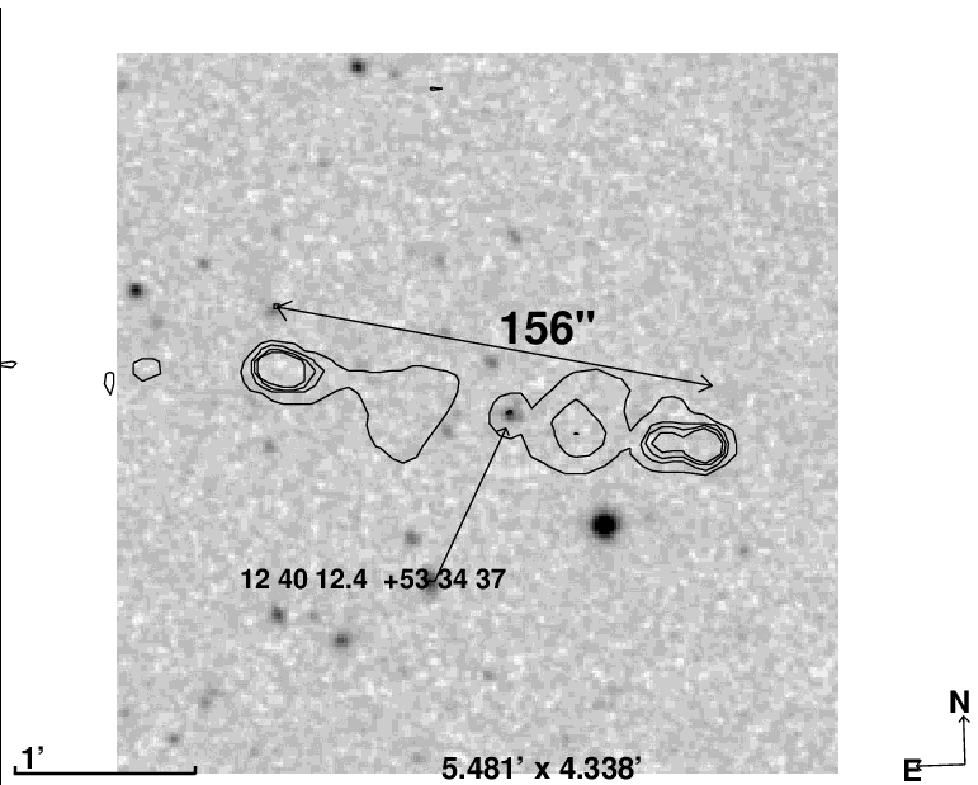}}
 \hspace{1mm}
\includegraphics[width=9.4cm,bb=0 15 285 215,clip]{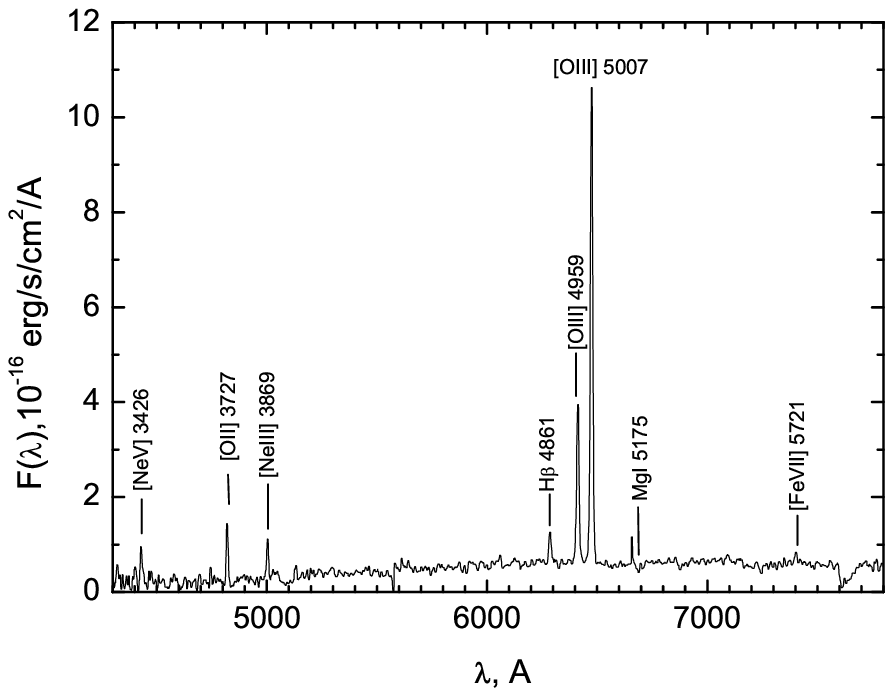}}
\caption{The FIRST  map contours. The rest is similar to
Fig.~\ref{fig1} for the source 1240+5334. }
\label{fig5}
\end{figure*}

\begin{figure*}
\centerline{
\hspace{4mm}\raisebox{9mm}{\includegraphics[width=7cm]{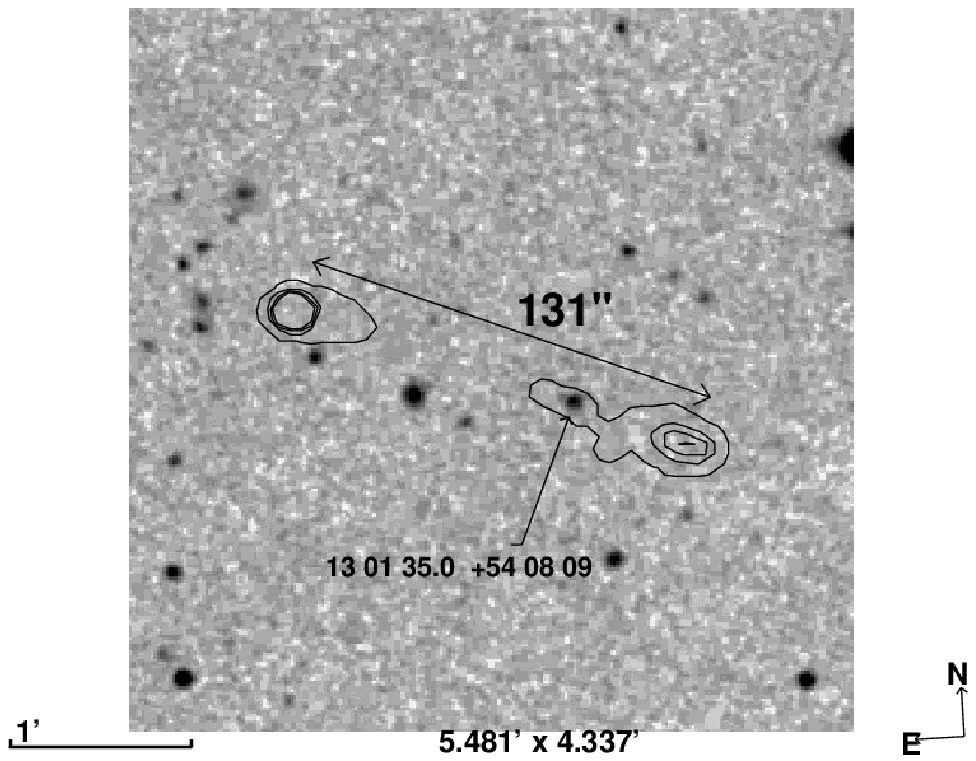}}
\includegraphics[width=9.4cm,bb=0 15 285 215,clip]{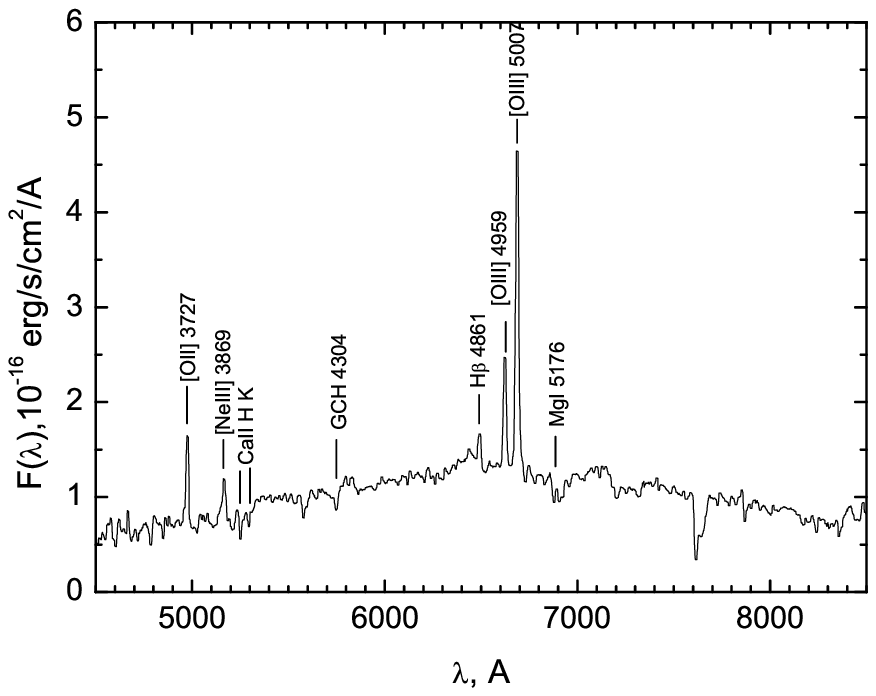}}
\caption{Similar to Fig.~\ref{fig5} for the source 1301+5408. }
\label{fig6}
\end{figure*}

\begin{figure*}
\centerline{
\hspace{1mm}\raisebox{9mm}{\includegraphics[width=7cm]{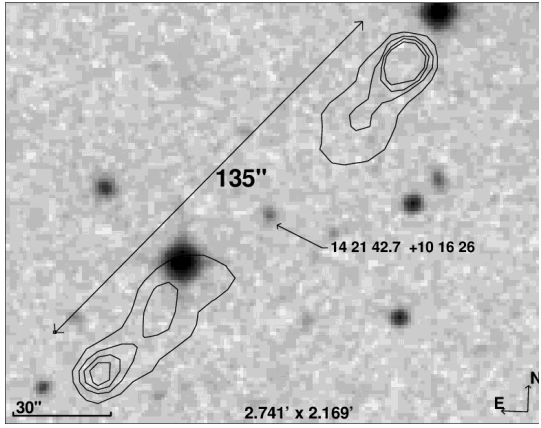}}
 \hspace{-4mm}
\includegraphics[width=9.4cm,bb=0 15 285 215,clip]{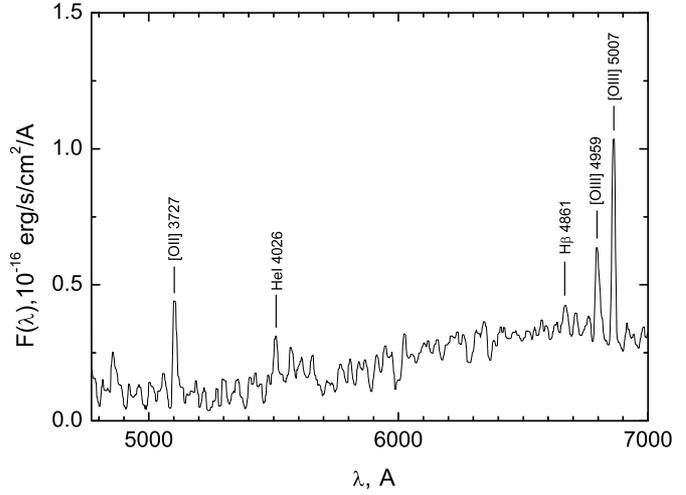}}
\caption{Similar to Fig.~\ref{fig5} for the source 1421+1016. }
\label{fig7}
\end{figure*}

\begin{figure*}
\centerline{
\hspace{2mm}\raisebox{9mm}{\includegraphics[width=7cm]{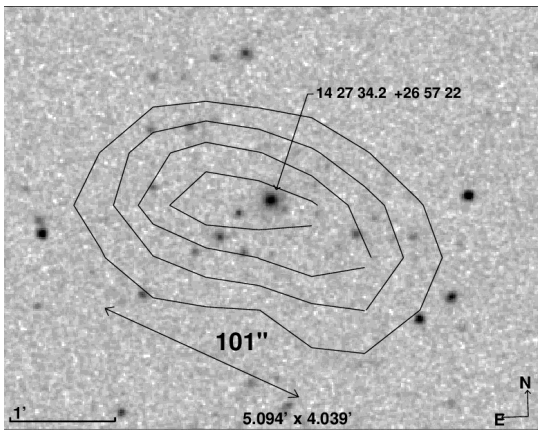}}
\includegraphics[width=9.4cm,bb=0 15 285 215,clip]{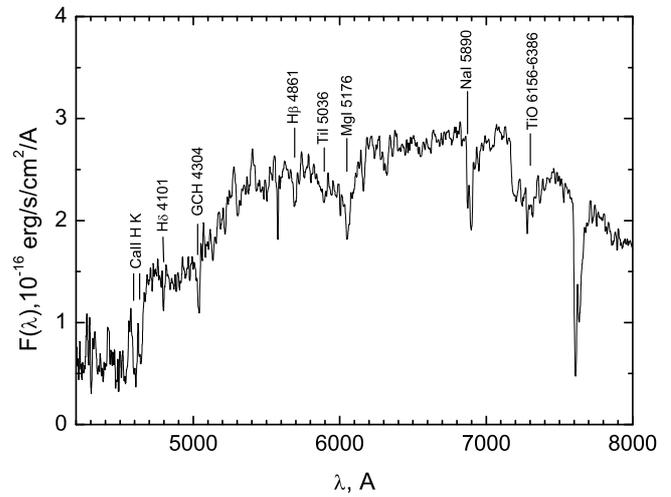}}
\caption{ Similar to Fig.~\ref{fig1} for the source
1427+2657. } \label{fig8}
\end{figure*}

\begin{figure*}
 \centerline{
\hspace{2mm}\includegraphics[width=7cm]{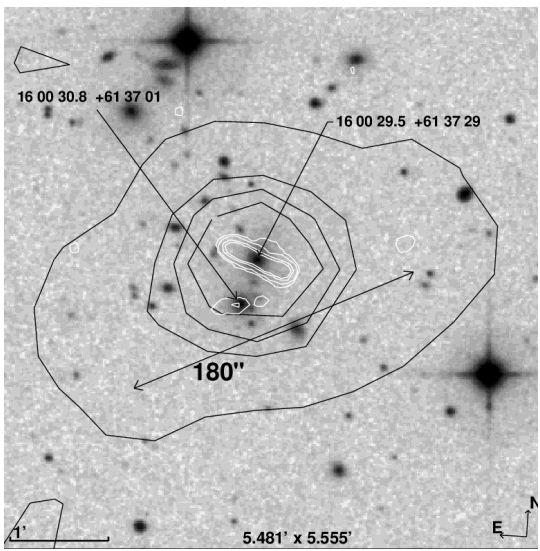}
\includegraphics[width=9.4cm,bb=0 15 285 215,clip]{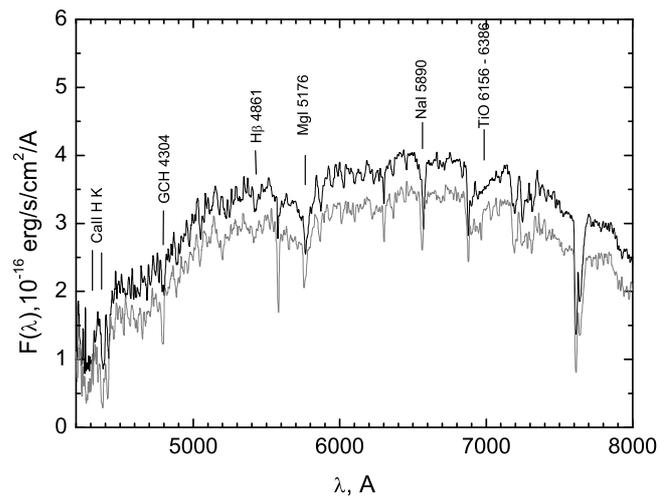}}
\caption{Similar to Fig.~\ref{fig1} for the source
1600+6137. White inner contours denote the FIRST data, the outer contours---the NVSS data. The spectra of both galaxies are shown. To distinguish between the spectra, the spectrum of the Nothern   galaxy is shifted upward along the vertical axis.}
\label{fig9}
\end{figure*}

\begin{table*}
\caption{The radio source parameters}
\label{tab2}
\medskip
\begin{tabular}{c|c|c|c|c|c|c}
\hline
Object name  &  RA 2000.0 &  DEC 2000.0 &  $\Theta$, arcsec  &  $S$, mJy   &   $z$ &    $D$, Mpc \\
\hline
  (1)     &     (2)    &     (3)     &     (4)     &   (5)   & (6)   & (7)        \\
\hline
0003+0351 & 00 03 32.83&  +03 51 14.2&  1174       & 450.8   & 0.097 &   2.08     \\
0422+1512 & 04 22 18.53&  +15 12 39.3&   735       & 168.8   & 0.409 &   3.98     \\
0638+3334 & 06 38 52.79&  +33 34 08.1&   201       & 268.7   & 0.088 &   0.33     \\
1124+7102 & 11 24 35.89&  +71 02 08.5&   164       &  83.3   & 0.141 &   0.40     \\
1240+5334 & 12 40 13.66&  +53 34 38.5&   156       & 276.8   & 0.293 &   0.68     \\
1301+5408 & 13 01 38.00&  +54 08 21.4&   131       & 276.4   & 0.336 &   0.62     \\
1421+1016 & 14 21 43.30&  +10 16 25.0&   135       & 238.8   & 0.370 &   0.69     \\
1427+2657 & 14 27 34.10&  +26 57 11.2&   101       &  45.3   & 0.170 &   0.29     \\
1600+6137 & 16 00 28.17&  +61 37 17.3&   180       & 129.7   & 0.114 &   0.35     \\
\hline
\end{tabular}
\end{table*}

\section{OBSERVATIONS}

The spectroscopic observations of the host galaxies were carried out at
the  6-m BTA telescope of the Special Astrophysical Observatory of the Russian Academy of Sciences with a multi-mode device SCORPIO-2 \citep{scorpio2}
as  an auxiliary program when the weather conditions prevented the implementation of the main program.  The observations were made in the long-slit spectroscopy mode, with the VPHG940@600 grism  providing the spectral
range of   3700--8500~\AA\, with the resolution of about 7~\AA\, at the slit width of   $1\arcsec$.

The data reduction was carried out by the authors' software  running in the  {\tt IDL} environment. The toolkit
performs conventional procedures in the present sequence: bias and  flat-field correction, wavelength calibration, flux calibration using
the spectra of the spectrophotometric standard stars observed in the same night. To extract the spectrum of the studied object with the maximum signal-to-noise ratio, the 2D-spectrum was convoluted with the weighted average cross-dispersion profile of the standard star
spectrum. The log of observations is given in
Table~\ref{tab1}.

\section{RESULTS}

Table~\ref{tab2}  lists the  observed
radio sources and their parameters: (1)~the name of the object;
(2,~3)~the coordinates of the radio source centroid; (4)~the apparent
angular size, measured as the angular distance between the most
distant components of the radio source; (5)~the integral flux
at the frequency of 1.4~GHz; (6)~redshift; (7)~the linear size
within   the  $\Lambda {\rm CDM}$ standard cosmological model.

Figures~\ref{fig1}--\ref{fig9}
present the NVSS and/or FIRST maps of radio sources, combined with the
 DSS2-red images. The maps show  the angular size of the
radio source (the long arrow parallel to the major axis of the
radio source),   position and coordinates of the host galaxy.
The right-hand side plots demonstrate the  spectra of the optical components of radio sources.

The field of the 1600+6137 source captured two galaxies
separated by  $28\arcsec$ (54~kpc in projection). The spectrograph slit
passes through both  objects. The resulting spectra are very similar, what is not
surprising for the elliptical galaxies  similar in size and
luminosity. The  radial velocity difference of these galaxies is
about \mbox {200--300~km\,s$^{-1}$};   they  most likely form a
physical pair. The radio images of these objects with high angular
resolution from the FIRST survey are shown in Fig.~9 by the white contours.
Since the angular resolution of the NVSS and FIRST is $45\arcsec$ and
$5\arcsec$, respectively, faint  extended  regions  detected by the NVSS
are not visible in the FIRST. At the same time, the activity of these
objects in the radio range according to the  FIRST is
one and a half order: the flux at 1.4~GHz from the  Northern galaxy at the
77.5~mJy, while it  is 2.63~mJy for the Southern galaxy.
The big difference between the fluxes might be due to the anisotropy of the radiation in the radio range and  different spatial orientations of the
radio source structures.

Table~2 contains the measured redshifts and calculated
apparent sizes of the observed radio sources. The calculations assumed
\mbox {$ \Omega_m=0.27$}, \mbox {$\Omega_v=0.73$}, \mbox {$H_0$ =71~km\,s$^{-1}$\,Mpc$^{-1}$}.
Two objects from our list got past the conditional boundary of
1~Mpc and hence make it to the list of giants. The most extended  object of the sample,  0422+1512
is only a little smaller  than the longest  known radio source  3C\,236.
\citet{Machalski} give the physical size of this object,
5.65~Mpc, assuming the Friedmann cosmology with the
Hubble constant of  \mbox {$H=50$~km\,s$^{-1}$\,Mpc$^{-1}$} and the
deceleration parameter  $q_0 = 0.5$.  In the current model, given
the angular size of 3C\,236  of around $2300\arcsec$, its  physical size
is 4.15~Mpc, which is practically the same  as that of  0422+1512.

\begin{acknowledgements}
This paper is based on the observations carried out at the 6-m BTA telescope of the Special Astrophysical Observatory of the RAS, operated with the financial support of the Ministry of Education and Science of the Russian Federation (agreement No.~14.619.21.0004, project ID RFMEFI61914X0004). A.V.M. is grateful for the financial support of the  ``Dynasty'' Foundation. This research has made use of the NASA/IPAC Extragalactic Database (NED) which is operated by the Jet Propulsion Laboratory, California Institute of Technology, under contract with the National Aeronautics and Space Administration.
\end{acknowledgements}

\end{document}